\documentclass[%
 aip,
 jmp,%
 amsmath,amssymb,
 preprint,%
]{revtex4-2}

\usepackage{graphicx}
\usepackage{dcolumn}
\usepackage{bm}

\begin{document}


\title{Continuous Varifocal Metalens Based on Phase-Change Material}

\author{Yilong Cui}
\affiliation{School of Information Engineering, Nanchang University, Nanchang 330031, China}

\author{Liang Hou}
\affiliation{School of Physics and Materials Science, Nanchang University, Nanchang 330031, China}

\author{Kenan Guo}
\affiliation{School of Information Engineering, Nanchang University, Nanchang 330031, China}

\author{Yue Jiang}
\affiliation{Jiluan Academy, Nanchang University, Nanchang 330031, China.}

\author{Qiegen Liu}
\affiliation{School of Information Engineering, Nanchang University, Nanchang 330031, China}

\author{Shuyuan Xiao}%
\email{syxiao@ncu.edu.cn}
\affiliation{School of Information Engineering, Nanchang University, Nanchang 330031, China}
\affiliation{Institute for Advanced Study, Nanchang University, Nanchang 330031, China}

\author{Tingting Liu}%
\email{ttliu@ncu.edu.cn}
\affiliation{School of Information Engineering, Nanchang University, Nanchang 330031, China}
\affiliation{Institute for Advanced Study, Nanchang University, Nanchang 330031, China}

\date{\today}

\begin{abstract}
	
Metasurfaces have provided new opportunities for the realization of flat lenses, among which tunable metalenses have garnered considerable attention due to their flexible functionalities. In this paper, we present a continuously tunable metalens based on the phase-change material Sb$_{2}$S$_{3}$, which enables precise and continuous focal length control through the transition of states. Under the excitation of linearly polarized light at 1550 nm, phase compensation is provided by changing the crystallization state of the Sb$_{2}$S$_{3}$ nanopillars, allowing the focal length to continuously shift between 36~$\mu$m and 48~$\mu$m. At the same time, the metalens maintains a high focusing efficiency over 75\%. This approach provides greater design flexibility and broader applicability across diverse applications. By reducing the reliance on polarized light sources, it enhances device integration and tunability, paving the way for new opportunities in the practical implementation of metalenses in advanced optical imaging and nanophotonics.

\end{abstract}

\maketitle

\section{Introduction}

optical lenses are essential components in optical systems such as microscopes, drones, and telescopes. Traditional lenses modify the phase of light by accumulating phase throughout the optical path, with their shape and material determining the phase of light. However, this method often suffers from disadvantages such as large size and limited functionality, making them difficult to align with the trends of miniaturization and integration. Metalenses, as an emerging technology based on metasurfaces, offer new possibilities for the miniaturization of optical lenses due to their compact size and unprecedented ability to control the incident light wavefront for focusing purposes\cite{Khorasaninejad2017,Tseng2018,Arbabi2022,Yao2023}. By adjusting parameters such as the shape, rotation direction, and height of the nanostructure, precise control over the polarization, phase, and amplitude of light can be achieved\cite{Fan2024,Li2021,Du2022,Kamali2018,Liu2024}. Furthermore, the functionality of metalens far exceeds that of traditional lenses, enabling applications such as achromatic lenses\cite{Fan2019,Wang2021,Hu2023}, optical imaging\cite{Yoon2021}, optical tweezers\cite{Li2021a}, and high-dimensional quantum sources\cite{Li2020,Li2023}. As a revolutionary technology in the field of optics, metalenses not only overcome the limitations of traditional optical systems but also have the potential to radically transform the complex lens designs in conventional optical systems, leading to the development of smaller, thinner, and lighter devices like smartphones, cameras, and surveillance cameras.

As the study of metalenses progresses, certain limitations have become apparent. For instance, the focal length of a metalens cannot be dynamically adjusted after fabrication, which restricts its potential for broader adoption in applications requiring dynamic tunability and reconfigurability\cite{Xiao2020}. To address this issue, some researchers have achieved tunable focal functionality by stretching an elastic substrate\cite{Ee2016}. However, this method becomes difficult to restore to its original state after large deformations or multiple uses. Some studies incorporate liquid crystal materials into metalenses, although this adds complexity to the system\cite{Bosch2021,Zhu2023}. Alternative methods for implementing varifocal functionality include adding extra optical elements, such as an output polarizer or diffractive optical elements (DOEs)\cite{Luo2021}. However, these approaches often lead to an increase in system size. In this context, using phase-change materials (PCMs) to realize actively tunable devices has become a promising approach. PCMs, such as vanadium dioxide (VO$_{2}$), antimony telluride (GST), and trisulfide (Sb$_{2}$S$_{3}$), exhibit significant differences in optical constants between different crystallization states, induced by thermal, optical, or electrical stimuli\cite{Ding2019}. Such active materials exhibit reversible, non-volatile, and ultrafast transitions between different states, which enables more flexible manipulation, making them highly valuable in the development of tunable metalens. In a straightforward manner, researchers categorized the PCM-based nanostructures into distinct regions or layers, achieving tunable bifocal metalenses by modulating the optical responses between the crystalline and amorphous states\cite{Qin2021,Li2024}. To prevent reduced compactness and increased fabrication complexity, recent studies have successfully achieved continuous focal adjustment of metalenses by combining the propagation phase and geometric phase within a single helical metasurface\cite{Zhang2024,Zhang2023}. However, existing metalenses that rely on circularly polarized light often require complex light source designs, which limit their flexibility and general applicability in practical scenarios. Therefore, it is imperative to develop metalenses that can overcome the dependence on circularly polarized light while enabling continuous focal control and maintaining high focusing efficiency.

In this work, we combine PCMs with metasurfaces to realize a tunable metalens consisting of Sb$_{2}$S$_{3}$ pillars and a SiO$_{2}$ substrate. Using the phase-change properties of Sb$_{2}$S$_{3}$, when its refractive index gradually changes from 2.9 to 3.3, the phase compensation provided by the material also gradually increases from 0 to its maximum value, causing the focal length to gradually increase as designed. We successfully demonstrate the metalens operating under linearly polarized light at a wavelength of 1550 nm, with the focal length continuously adjustable between 36~$\mu$m and 48~$\mu$m. At the same time, during the focal length variation, the metalens maintains a high focusing efficiency (always above 75\%). Our design relies entirely on propagation phase control, reducing the need for specific polarized light sources and simplifying the design process of metalenses. It ensures high efficiency and functionality while expanding application scenarios, offering new ideas for diverse optical device designs.

\section{Results and Discussion}

The metalens consists of Sb$_{2}$S$_{3}$ nanopillars with a length of \textit{L} and a height of \textit{H}, positioned atop a glass substrate. Since PCMs exhibit varying optical constants in different crystallization states\cite{Lu2021}, the crystallization state of the Sb$_{2}$S$_{3}$ can be adjusted to meet the phase requirements of the metalens at different focal lengths, enabling the possibility of achieving continuous focal adjustment. As shown in Fig. 1, as the refractive index of the Sb$_{2}$S$_{3}$ nanopillars increases, the focal length of the metalens gradually shifts.

\begin{figure*}[htbp]
	\centering
	\includegraphics
	[scale=0.3]{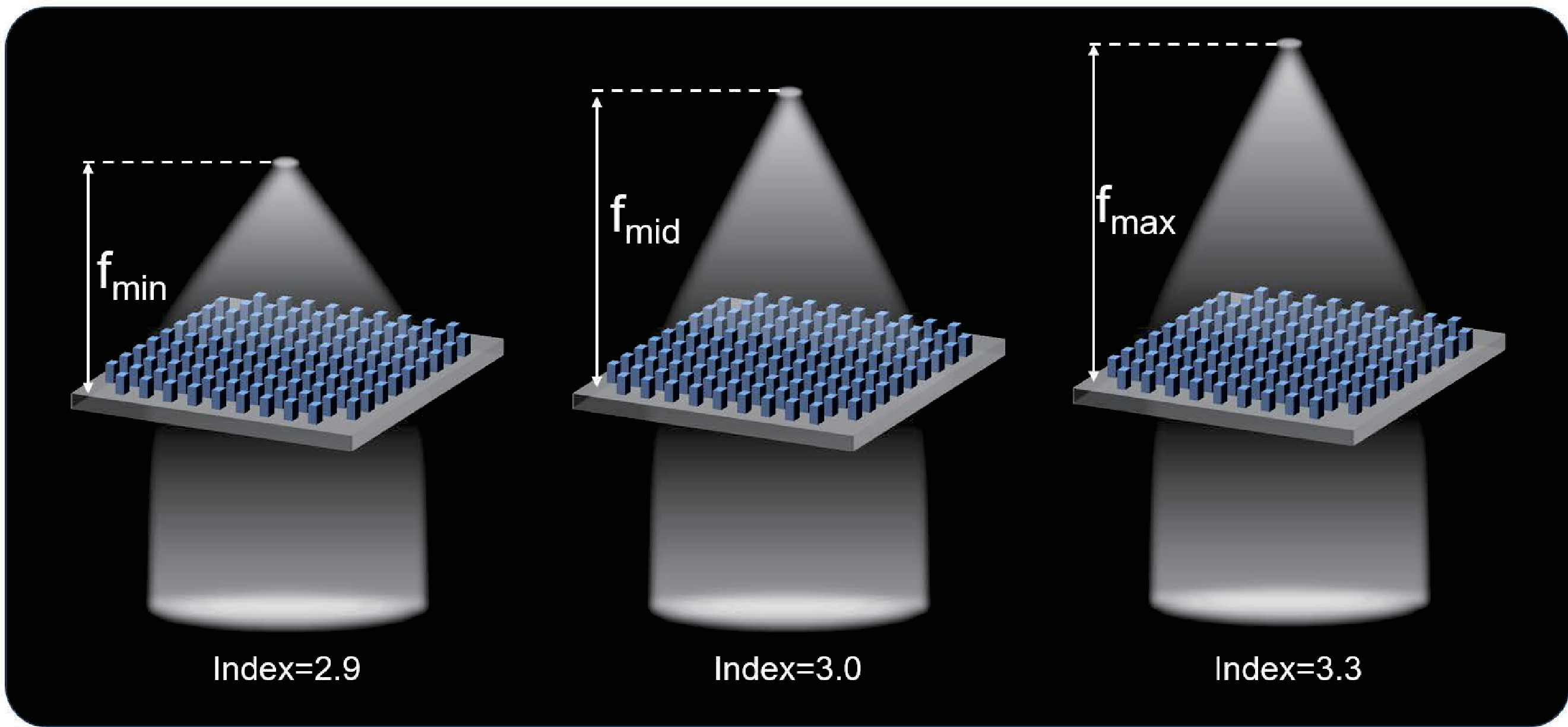}
	\caption{\label{Fig1} Schematic of the tunable metalens. The focal length can be varifocally tuned by varying the refractive index of the Sb$_{2}$S$_{3}$ nanopillars. The inset shows the schematic of the metalens at refractive indices of 2.9, 3.3, and intermediate values.}
\end{figure*}

When the focal length is fixed, the phase distribution of a metalens can be expressed as\cite{Li2024a}
\begin{equation}
	\label{deqn_ex1a}
	\phi(x, y) = -\frac{2\pi}{\lambda} \left( \sqrt{x^2 + y^2 + f^2} - f \right),
\end{equation}
where $\phi$ represents the phase of the metalens, $x$ and $y$ are the spatial coordinates in the $x$ and $y$ directions, $\lambda$ is the operating wavelength, and $f$ is the focal length of the metalens. To achieve the varifocal effect, we need to provide the required phase for the metalens at different focal lengths. To this end, we consider the phase compensation provided by the Sb$_{2}$S$_{3}$ during the refractive index change, and relate the focal length $f$, phase $\phi$, and refractive index $n$ in the phase formula. Therefore, the phase formula can be re-expressed as
\begin{equation}
	\phi(n, x, y) = -\frac{2\pi}{\lambda} \left( \sqrt{x^2 + y^2 + f(n)^2} - f(n) \right),
	\label{eq:phase_formula}
\end{equation}

Then we divide the phase $\phi$ in Eq.~(2) into a base phase $\phi(x, y, n_{\text{min}})$ and a phase compensation $\Delta\phi(n)$ that satisfies the phase requirements during the focal length variation. The $\phi(n, x, y)$ in Eq.~(2) can be written as the superposition of a fixed phase component and a variable phase component as
\begin{equation}
	\phi(n, x, y) = \phi(x, y, n_{\text{min}}) + \Delta\phi(n),
	\label{eq:phase_superposition}
\end{equation}
where $\phi(n, x, y)$ represents the total phase of the metalens, $\phi(x, y, n_{\text{min}})$ is the base phase provided by the Sb$_2$S$_3$ nanopillars when the refractive index is 2.9, and $\Delta\phi(n)$ represents the phase compensation with additional propagation phase that needs to be specifically designed. To achieve the continuously tunable functionality, it is necessary to provide the corresponding phase compensation to meet the phase requirements of the metalens at different focal lengths.

\begin{figure}[htbp]
	\centering
	\includegraphics
	[scale=0.35]{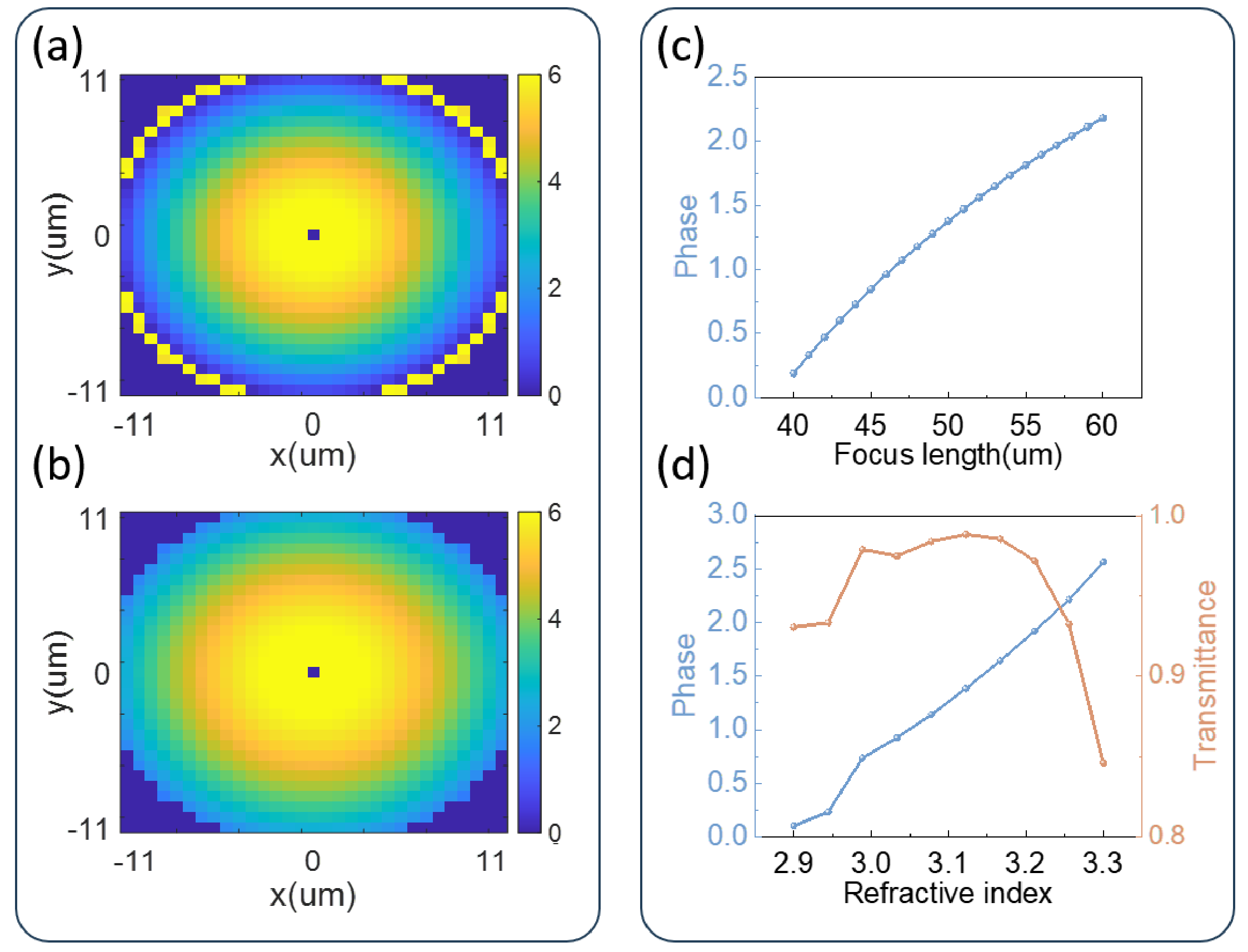}
	\caption{\label{Fig2} Phase distribution of the metalens at specific focal lengths and its relationship with the phase variation of unit cells. (a) and (b) show the phase distributions of the metalens at focal lengths of 40~$\mu$m and 60~$\mu$m, respectively. (c) Phase variation of the metalens at a fixed position as the focal length changes, derived based on Equation~(3). (d) Illustration of the phase and transmittance of the selected unit cell as a function of refractive index.}
\end{figure}

To more accurately determine the relationship between the metalens phase and focal length variation, we analyze and calculate the phase change under two different fixed focal lengths. Figs. 2(a) and 2(b) show the phase profiles for the metalens working at focal lengths of 40~$\mu$m and 60~$\mu$m, respectively. It is evident that as the focal length increases, the overall phase of the metalens also increases. For a clearer insight, we investigate the dependence of the phase $\phi$ on the focal length $f$ at a specific wavelength, while keeping the plane coordinates fixed. In this context, Eq.~(1) can be expressed as
\begin{equation}
	\phi(f) = -\frac{2\pi}{\lambda} \left( \sqrt{r^2 + f^2} - f \right),
	\label{eq:phase_focal_length}
\end{equation}
\noindent
where $r^2=x^2+y^2$ is considered as a fixed value. Based on the equation, we observe that the phase $\phi$ shows a monotonic increase as the focal length $f$ increases, as shown in Fig. 2(c). This suggests that the key to designing a continuously varifocal metalens lies in constructing a library of Sb$_{2}$S$_{3}$ nanopillars capable of matching the required phase compensation variations. Accordingly, different nanopillars with precise phase shifts must be engineered. Specifically, as the refractive index increases from 2.9 to 3.3, these nanopillars should provide the necessary phase compensation for the metalens, as illustrated in Fig. 2(d).

\begin{figure}[htbp]
	\centering
	\includegraphics
	[scale=0.22]{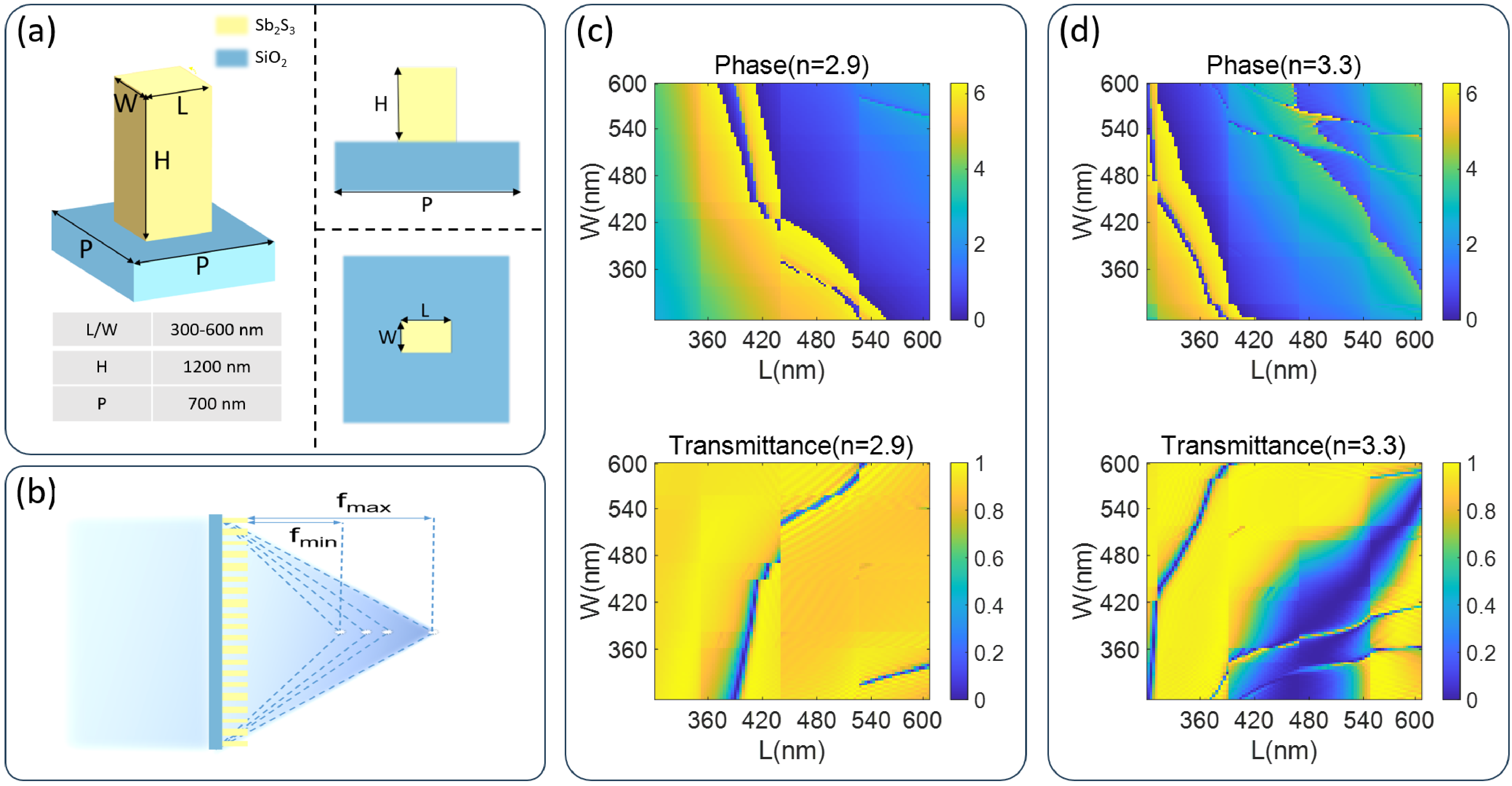}
	\caption{\label{Fig3}  Design of tunable metalenses. (a) Schematic of the metalens unit cell, where the nanopillar width (\textit{W}) or length (\textit{L}) ranges from 300 to 600~nm, height (\textit{H}) is 1200~nm, and structural period (\textit{P}) is 700~nm. (b) Schematic representation of the tunable metalens. (c) and (d) Phase and transmission distribution profiles of the nanopillars for refractive indices of 2.9 and 3.3, corresponding to different \textit{W} or \textit{L} values.}
\end{figure}

In this work, the Finite Difference Time Domain (FDTD) method is used to accurately calculate the response of the unit cell. Periodic boundary conditions are applied in the $x$ and $y$ directions, while Perfectly Matched Layer (PML) is used in the $z$ direction. Fig. 3(a) shows the unit cell selected for this study, where the period $P$ of the SiO$_2$ substrate is 700~nm, and the height $H$ of the Sb$_{2}$S$_{3}$ nanopillars is 1200~nm. By scanning the length ($L$) and width ($W$) of the unit cell, phase and transmission profiles of the Sb$_{2}$S$_{3}$ structure at refractive indices $n$ of 2.9 and 3.3 are calculated, as shown in Figs. 3(c) and 3(d). The unit cells with transmission within a relatively large value are considered. As the metalens in this study is designed for linearly polarized light, its phase is entirely governed by the propagation phase. Therefore, in selecting the structure, we focus solely on the propagation phase provided by each unit cell. After calculating the required phase distributions for both maximum and minimum focal lengths, we designate the phase of the nanopillars at $n=2.9$ as the base phase for the metalens at minimum focal length. By utilizing the additional propagation phase induced by changes in refractive index as phase compensation, we can adjust the focal length precisely, thereby defining the dimensions of each nanopillar, as shown in Table~1.

\begin{figure}[htbp]
	\centering
	\includegraphics
	[scale=0.3]{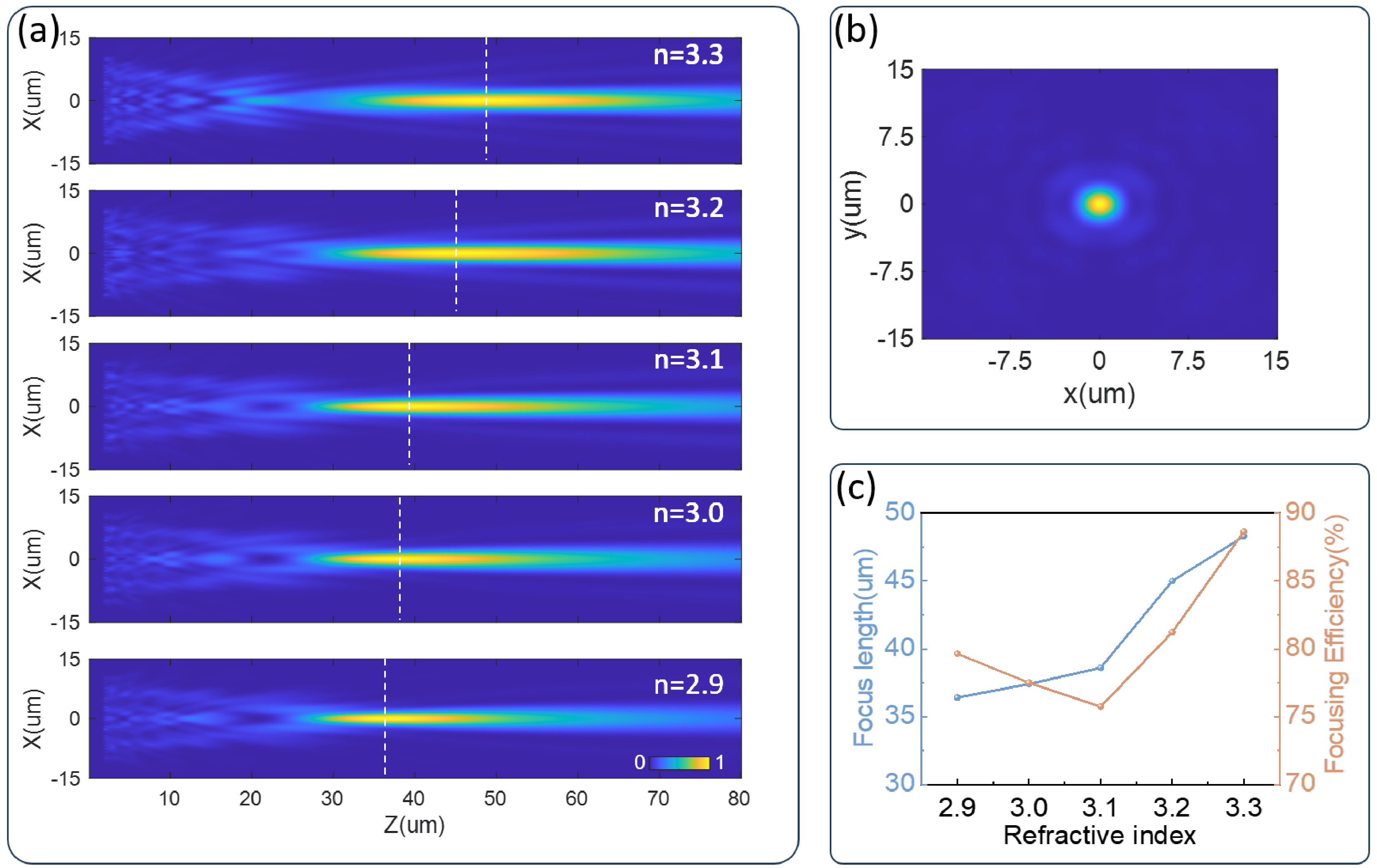}
	\caption{\label{Fig4} Simulation results of the metalens. (a) Focusing intensity distribution of the metalens in the \textit{x-z} plane under different refractive indices. (b) Intensity distribution of the metalens in the \textit{x-y} plane. (c) Line graph depicting the variation in focal length and focusing efficiency of the metalens as the refractive index increases. }
\end{figure}

The selected nanostructures with specific phase and phase differences are arranged as the layout of the metalens, and the proposed functionality of tunable focal lengths can thus be successfully achieved. Fig. 4 presents the full-wave simulation results obtained using the FDTD method. As shown in Fig. 4(a) for the intensity distribution of the metalens in the \textit{x-z} plane, it can be clearly observed a gradual change of the focal length from 36~$\mu$m to 48~$\mu$m. The difference between the expected and simulated focal lengths mainly arises from the small size of the metalens in simulation. Fig. 4(b) presents the intensity distribution of the metalens in the \textit{x-y} plane. To visually observe the trend of focal length variation, we also plot the relationship between focal length and focusing efficiency of the metalens as a function of Sb$_{2}$S$_{3}$ refractive index, as shown in Fig. 4(c). As the refractive index increases, the variation rate of focal length gradually becomes more pronounced, which is closely associated with the optical properties of Sb$_{2}$S$_{3}$. It is noted that the phase change of Sb$_{2}$S$_{3}$ intensifies with increasing refractive index, whereas the phase change of the metalens diminishes as the focal length increases, resulting in non-uniform focal length variation. Throughout the focal length adjustment process, the metalens maintains a high focusing efficiency, consistently exceeding 75\%.

\begin{table*}[htbp]
	\caption{Selected unit cell parameters used in the article.}
	\centering
	\scalebox{1}{
		\begin{tabular}{ccccccccccccccc}
			\hline
			\textbf{Structure} & \textbf{1} & \textbf{2} & \textbf{3} & \textbf{4} & \textbf{5} & \textbf{6} & \textbf{7} & \textbf{8} & \textbf{9} & \textbf{10} & \textbf{11} & \textbf{12} & \textbf{13} & \textbf{14}\\ 
			\hline
			\textit{L} [nm] & 600 & 408 & 300 & 300 & 393 & 447 & 585 & 453 & 321 & 300 & 309 & 411 & 300 & 420 \\ 
			\textit{W} [nm] & 600 & 558 & 357 & 486 & 600 & 576 & 576 & 537 & 600 & 405 & 384 & 600 & 546 & 597\\ 
			$n=2.9$ Phase [rad] & 2.53 & 6.05 & 2.84 & 3.49 & 5.55 & 0.70 & 2.06 & 0.58 & 4.45 & 3.06 & 3.10 & 0.10 & 3.72 & 0.26\\ 
			$n=3.3$ Phase [rad] & 4.04 & 2.45 & 4.50 & 5.94 & 2.54 & 2.99 & 3.66 & 3.24 & 0.29 & 4.85 & 5.01 & 2.88 & 0.10 & 2.95\\ 
			$n=2.9$ Transmittance [\%] & 85 & 97 & 93 & 93 & 76 & 88 & 93 & 88 & 99 & 91 & 91 & 98 & 93 & 95\\ 
			$n=3.3$ Transmittance [\%] & 85 & 50 & 97 & 84 & 32 & 10 & 93 & 16 & 97 & 98 & 98 & 32 & 58 & 20\\ 
			\hline
		\end{tabular}
	}
	\label{tab:merged_table}
\end{table*}

\section{Conclusions}

In conclusion, we propose a tunable metalens based on the Sb$_{2}$S$_{3}$. As the metalens focal length increases, its phase requirements also gradually increase. By providing the appropriate phase compensation, continuous focal length adjustment can be achieved. This study employs the tunable optical constants of Sb$_{2}$S$_{3}$ in various crystalline phases to facilitate phase compensation. This allows for continuous focal length tunability of the metalens from 36~$\mu$m to 48~$\mu$m under 1550 nm linearly polarized light. The designed metalens maintains a focusing efficiency above 75\% throughout the zooming process. The metalens we designed is excited by linearly polarized light, eliminating the need for a quarter-wave plate. This simplification improves both integration and tunability, opening new possibilities for high-efficiency optical imaging and nanophotonics applications.

\section*{Data availability}

Relevant data supporting the key findings of this study are available within the article. All raw data generated during the current study are available from the corresponding authors upon reasonable request.

\section*{Conflicts of interest}

The authors declare no conflict of interest.

\begin{acknowledgments}
	
This work was supported by the National Natural Science Foundation of China (Grants No. 12364045, No. 12264028, and No. 12304420), the Natural Science Foundation of Jiangxi Province (Grants No. 20232BAB201040 and No. 20232BAB211025), and the Young Elite Scientists Sponsorship Program by JXAST (Grants No. 2023QT11 and 2025QT04). 

\end{acknowledgments}



\nocite{*}

\begin{thebibliography}{28}%
	\makeatletter
	\providecommand \@ifxundefined [1]{%
		\@ifx{#1\undefined}
	}%
	\providecommand \@ifnum [1]{%
		\ifnum #1\expandafter \@firstoftwo
		\else \expandafter \@secondoftwo
		\fi
	}%
	\providecommand \@ifx [1]{%
		\ifx #1\expandafter \@firstoftwo
		\else \expandafter \@secondoftwo
		\fi
	}%
	\providecommand \natexlab [1]{#1}%
	\providecommand \enquote  [1]{``#1''}%
	\providecommand \bibnamefont  [1]{#1}%
	\providecommand \bibfnamefont [1]{#1}%
	\providecommand \citenamefont [1]{#1}%
	\providecommand \href@noop [0]{\@secondoftwo}%
	\providecommand \href [0]{\begingroup \@sanitize@url \@href}%
	\providecommand \@href[1]{\@@startlink{#1}\@@href}%
	\providecommand \@@href[1]{\endgroup#1\@@endlink}%
	\providecommand \@sanitize@url [0]{\catcode `\\12\catcode `\$12\catcode
		`\&12\catcode `\#12\catcode `\^12\catcode `\_12\catcode `\%12\relax}%
	\providecommand \@@startlink[1]{}%
	\providecommand \@@endlink[0]{}%
	\providecommand \url  [0]{\begingroup\@sanitize@url \@url }%
	\providecommand \@url [1]{\endgroup\@href {#1}{\urlprefix }}%
	\providecommand \urlprefix  [0]{URL }%
	\providecommand \Eprint [0]{\href }%
	\providecommand \doibase [0]{https://doi.org/}%
	\providecommand \selectlanguage [0]{\@gobble}%
	\providecommand \bibinfo  [0]{\@secondoftwo}%
	\providecommand \bibfield  [0]{\@secondoftwo}%
	\providecommand \translation [1]{[#1]}%
	\providecommand \BibitemOpen [0]{}%
	\providecommand \bibitemStop [0]{}%
	\providecommand \bibitemNoStop [0]{.\EOS\space}%
	\providecommand \EOS [0]{\spacefactor3000\relax}%
	\providecommand \BibitemShut  [1]{\csname bibitem#1\endcsname}%
	\let\auto@bib@innerbib\@empty
	\bibitem [{\citenamefont {Khorasaninejad}\ and\ \citenamefont
		{Capasso}(2017)}]{Khorasaninejad2017}%
	\BibitemOpen
	\bibfield  {author} {\bibinfo {author} {\bibfnamefont {M.}~\bibnamefont
			{Khorasaninejad}}\ and\ \bibinfo {author} {\bibfnamefont {F.}~\bibnamefont
			{Capasso}},\ }\bibfield  {title} {\enquote {\bibinfo {title} {Metalenses:
				Versatile multifunctional photonic components},}\ }\href
	{https://doi.org/10.1126/science.aam8100} {\bibfield  {journal} {\bibinfo
			{journal} {Science}\ }\textbf {\bibinfo {volume} {358}},\ \bibinfo {pages}
		{eaam8100} (\bibinfo {year} {2017})}\BibitemShut {NoStop}%
	\bibitem [{\citenamefont {Tseng}\ \emph {et~al.}(2018)\citenamefont {Tseng},
		\citenamefont {Hsiao}, \citenamefont {Chu}, \citenamefont {Chen},
		\citenamefont {Sun}, \citenamefont {Liu},\ and\ \citenamefont
		{Tsai}}]{Tseng2018}%
	\BibitemOpen
	\bibfield  {author} {\bibinfo {author} {\bibfnamefont {M.~L.}\ \bibnamefont
			{Tseng}}, \bibinfo {author} {\bibfnamefont {H.}~\bibnamefont {Hsiao}},
		\bibinfo {author} {\bibfnamefont {C.~H.}\ \bibnamefont {Chu}}, \bibinfo
		{author} {\bibfnamefont {M.~K.}\ \bibnamefont {Chen}}, \bibinfo {author}
		{\bibfnamefont {G.}~\bibnamefont {Sun}}, \bibinfo {author} {\bibfnamefont
			{A.}~\bibnamefont {Liu}},\ and\ \bibinfo {author} {\bibfnamefont {D.~P.}\
			\bibnamefont {Tsai}},\ }\bibfield  {title} {\enquote {\bibinfo {title}
			{Metalenses: Advances and applications},}\ }\href
	{https://doi.org/10.1002/adom.201800554} {\bibfield  {journal} {\bibinfo
			{journal} {Advanced Optical Materials}\ }\textbf {\bibinfo {volume} {6}},\
		\bibinfo {pages} {1800554} (\bibinfo {year} {2018})}\BibitemShut {NoStop}%
	\bibitem [{\citenamefont {Arbabi}\ and\ \citenamefont
		{Faraon}(2022)}]{Arbabi2022}%
	\BibitemOpen
	\bibfield  {author} {\bibinfo {author} {\bibfnamefont {A.}~\bibnamefont
			{Arbabi}}\ and\ \bibinfo {author} {\bibfnamefont {A.}~\bibnamefont
			{Faraon}},\ }\bibfield  {title} {\enquote {\bibinfo {title} {Advances in
				optical metalenses},}\ }\href {https://doi.org/10.1038/s41566-022-01108-6}
	{\bibfield  {journal} {\bibinfo  {journal} {Nature Photonics}\ }\textbf
		{\bibinfo {volume} {17}},\ \bibinfo {pages} {16--25} (\bibinfo {year}
		{2022})}\BibitemShut {NoStop}%
	\bibitem [{\citenamefont {Yao}\ \emph {et~al.}(2023)\citenamefont {Yao},
		\citenamefont {Lin}, \citenamefont {Chen},\ and\ \citenamefont
		{Tsai}}]{Yao2023}%
	\BibitemOpen
	\bibfield  {author} {\bibinfo {author} {\bibfnamefont {J.}~\bibnamefont
			{Yao}}, \bibinfo {author} {\bibfnamefont {R.}~\bibnamefont {Lin}}, \bibinfo
		{author} {\bibfnamefont {M.~K.}\ \bibnamefont {Chen}},\ and\ \bibinfo
		{author} {\bibfnamefont {D.~P.}\ \bibnamefont {Tsai}},\ }\bibfield  {title}
	{\enquote {\bibinfo {title} {Integrated-resonant metadevices: a review},}\
	}\href {https://doi.org/10.1117/1.ap.5.2.024001} {\bibfield  {journal}
		{\bibinfo  {journal} {Advanced Photonics}\ }\textbf {\bibinfo {volume} {5}},\
		\bibinfo {pages} {024001} (\bibinfo {year} {2023})}\BibitemShut {NoStop}%
	\bibitem [{\citenamefont {Fan}\ \emph {et~al.}(2024)\citenamefont {Fan},
		\citenamefont {Liang}, \citenamefont {Wang}, \citenamefont {Chen},
		\citenamefont {Lai}, \citenamefont {Ku~Chen}, \citenamefont {Xiao},
		\citenamefont {Li},\ and\ \citenamefont {Tsai}}]{Fan2024}%
	\BibitemOpen
	\bibfield  {author} {\bibinfo {author} {\bibfnamefont {Y.}~\bibnamefont
			{Fan}}, \bibinfo {author} {\bibfnamefont {H.}~\bibnamefont {Liang}}, \bibinfo
		{author} {\bibfnamefont {Y.}~\bibnamefont {Wang}}, \bibinfo {author}
		{\bibfnamefont {S.}~\bibnamefont {Chen}}, \bibinfo {author} {\bibfnamefont
			{F.}~\bibnamefont {Lai}}, \bibinfo {author} {\bibfnamefont {M.}~\bibnamefont
			{Ku~Chen}}, \bibinfo {author} {\bibfnamefont {S.}~\bibnamefont {Xiao}},
		\bibinfo {author} {\bibfnamefont {J.}~\bibnamefont {Li}},\ and\ \bibinfo
		{author} {\bibfnamefont {D.~P.}\ \bibnamefont {Tsai}},\ }\bibfield  {title}
	{\enquote {\bibinfo {title} {Dual-channel quantum meta-hologram for
				display},}\ }\href {https://doi.org/10.1117/1.apn.3.1.016011} {\bibfield
		{journal} {\bibinfo  {journal} {Advanced Photonics Nexus}\ }\textbf {\bibinfo
			{volume} {3}},\ \bibinfo {pages} {016011} (\bibinfo {year}
		{2024})}\BibitemShut {NoStop}%
	\bibitem [{\citenamefont {Li}\ \emph {et~al.}(2021{\natexlab{a}})\citenamefont
		{Li}, \citenamefont {Li}, \citenamefont {Yan}, \citenamefont {Xu},
		\citenamefont {Wang}, \citenamefont {Yao}, \citenamefont {Wang},\ and\
		\citenamefont {Zhu}}]{Li2021}%
	\BibitemOpen
	\bibfield  {author} {\bibinfo {author} {\bibfnamefont {T.}~\bibnamefont
			{Li}}, \bibinfo {author} {\bibfnamefont {X.}~\bibnamefont {Li}}, \bibinfo
		{author} {\bibfnamefont {S.}~\bibnamefont {Yan}}, \bibinfo {author}
		{\bibfnamefont {X.}~\bibnamefont {Xu}}, \bibinfo {author} {\bibfnamefont
			{S.}~\bibnamefont {Wang}}, \bibinfo {author} {\bibfnamefont {B.}~\bibnamefont
			{Yao}}, \bibinfo {author} {\bibfnamefont {Z.}~\bibnamefont {Wang}},\ and\
		\bibinfo {author} {\bibfnamefont {S.}~\bibnamefont {Zhu}},\ }\bibfield
	{title} {\enquote {\bibinfo {title} {Generation and conversion dynamics of
				dual bessel beams with a photonic spin-dependent dielectric metasurface},}\
	}\href {https://doi.org/10.1103/physrevapplied.15.014059} {\bibfield
		{journal} {\bibinfo  {journal} {Physical Review Applied}\ }\textbf {\bibinfo
			{volume} {15}},\ \bibinfo {pages} {014059} (\bibinfo {year}
		{2021}{\natexlab{a}})}\BibitemShut {NoStop}%
	\bibitem [{\citenamefont {Du}\ \emph {et~al.}(2022)\citenamefont {Du},
		\citenamefont {Barkaoui}, \citenamefont {Zhang}, \citenamefont {Jin},
		\citenamefont {Song},\ and\ \citenamefont {Xiao}}]{Du2022}%
	\BibitemOpen
	\bibfield  {author} {\bibinfo {author} {\bibfnamefont {K.}~\bibnamefont
			{Du}}, \bibinfo {author} {\bibfnamefont {H.}~\bibnamefont {Barkaoui}},
		\bibinfo {author} {\bibfnamefont {X.}~\bibnamefont {Zhang}}, \bibinfo
		{author} {\bibfnamefont {L.}~\bibnamefont {Jin}}, \bibinfo {author}
		{\bibfnamefont {Q.}~\bibnamefont {Song}},\ and\ \bibinfo {author}
		{\bibfnamefont {S.}~\bibnamefont {Xiao}},\ }\bibfield  {title} {\enquote
		{\bibinfo {title} {Optical metasurfaces towards multifunctionality and
				tunability},}\ }\href {https://doi.org/10.1515/nanoph-2021-0684} {\bibfield
		{journal} {\bibinfo  {journal} {Nanophotonics}\ }\textbf {\bibinfo {volume}
			{11}},\ \bibinfo {pages} {1761--1781} (\bibinfo {year} {2022})}\BibitemShut
	{NoStop}%
	\bibitem [{\citenamefont {Kamali}\ \emph {et~al.}(2018)\citenamefont {Kamali},
		\citenamefont {Arbabi}, \citenamefont {Arbabi},\ and\ \citenamefont
		{Faraon}}]{Kamali2018}%
	\BibitemOpen
	\bibfield  {author} {\bibinfo {author} {\bibfnamefont {S.~M.}\ \bibnamefont
			{Kamali}}, \bibinfo {author} {\bibfnamefont {E.}~\bibnamefont {Arbabi}},
		\bibinfo {author} {\bibfnamefont {A.}~\bibnamefont {Arbabi}},\ and\ \bibinfo
		{author} {\bibfnamefont {A.}~\bibnamefont {Faraon}},\ }\bibfield  {title}
	{\enquote {\bibinfo {title} {A review of dielectric optical metasurfaces for
				wavefront control},}\ }\href {https://doi.org/10.1515/nanoph-2017-0129}
	{\bibfield  {journal} {\bibinfo  {journal} {Nanophotonics}\ }\textbf
		{\bibinfo {volume} {7}},\ \bibinfo {pages} {1041--1068} (\bibinfo {year}
		{2018})}\BibitemShut {NoStop}%
	\bibitem [{\citenamefont {Liu}\ \emph {et~al.}(2024)\citenamefont {Liu},
		\citenamefont {Qiu}, \citenamefont {Xu}, \citenamefont {Qin}, \citenamefont
		{Wan}, \citenamefont {Yu}, \citenamefont {Liu}, \citenamefont {Huang},\ and\
		\citenamefont {Xiao}}]{Liu2024}%
	\BibitemOpen
	\bibfield  {author} {\bibinfo {author} {\bibfnamefont {T.}~\bibnamefont
			{Liu}}, \bibinfo {author} {\bibfnamefont {J.}~\bibnamefont {Qiu}}, \bibinfo
		{author} {\bibfnamefont {L.}~\bibnamefont {Xu}}, \bibinfo {author}
		{\bibfnamefont {M.}~\bibnamefont {Qin}}, \bibinfo {author} {\bibfnamefont
			{L.}~\bibnamefont {Wan}}, \bibinfo {author} {\bibfnamefont {T.}~\bibnamefont
			{Yu}}, \bibinfo {author} {\bibfnamefont {Q.}~\bibnamefont {Liu}}, \bibinfo
		{author} {\bibfnamefont {L.}~\bibnamefont {Huang}},\ and\ \bibinfo {author}
		{\bibfnamefont {S.}~\bibnamefont {Xiao}},\ }\bibfield  {title} {\enquote
		{\bibinfo {title} {Edge detection imaging by quasi-bound states in the
				continuum},}\ }\href {https://doi.org/10.1021/acs.nanolett.4c04543}
	{\bibfield  {journal} {\bibinfo  {journal} {Nano Letters}\ }\textbf {\bibinfo
			{volume} {24}},\ \bibinfo {pages} {14466--14474} (\bibinfo {year}
		{2024})}\BibitemShut {NoStop}%
	\bibitem [{\citenamefont {Fan}\ \emph {et~al.}(2019)\citenamefont {Fan},
		\citenamefont {Qiu}, \citenamefont {Zhang}, \citenamefont {Pang},
		\citenamefont {Zhou}, \citenamefont {Liu}, \citenamefont {Ren}, \citenamefont
		{Wang},\ and\ \citenamefont {Dong}}]{Fan2019}%
	\BibitemOpen
	\bibfield  {author} {\bibinfo {author} {\bibfnamefont {Z.-B.}\ \bibnamefont
			{Fan}}, \bibinfo {author} {\bibfnamefont {H.-Y.}\ \bibnamefont {Qiu}},
		\bibinfo {author} {\bibfnamefont {H.-L.}\ \bibnamefont {Zhang}}, \bibinfo
		{author} {\bibfnamefont {X.-N.}\ \bibnamefont {Pang}}, \bibinfo {author}
		{\bibfnamefont {L.-D.}\ \bibnamefont {Zhou}}, \bibinfo {author}
		{\bibfnamefont {L.}~\bibnamefont {Liu}}, \bibinfo {author} {\bibfnamefont
			{H.}~\bibnamefont {Ren}}, \bibinfo {author} {\bibfnamefont {Q.-H.}\
			\bibnamefont {Wang}},\ and\ \bibinfo {author} {\bibfnamefont {J.-W.}\
			\bibnamefont {Dong}},\ }\bibfield  {title} {\enquote {\bibinfo {title} {A
				broadband achromatic metalens array for integral imaging in the visible},}\
	}\href {https://doi.org/10.1038/s41377-019-0178-2} {\bibfield  {journal}
		{\bibinfo  {journal} {Light: Science \& Applications}\ }\textbf {\bibinfo
			{volume} {8}},\ \bibinfo {pages} {67} (\bibinfo {year} {2019})}\BibitemShut
	{NoStop}%
	\bibitem [{\citenamefont {Wang}\ \emph {et~al.}(2021)\citenamefont {Wang},
		\citenamefont {Chen}, \citenamefont {Yang}, \citenamefont {Ji}, \citenamefont
		{Jin}, \citenamefont {Ma}, \citenamefont {Song}, \citenamefont {Boltasseva},
		\citenamefont {Han}, \citenamefont {Shalaev},\ and\ \citenamefont
		{Xiao}}]{Wang2021}%
	\BibitemOpen
	\bibfield  {author} {\bibinfo {author} {\bibfnamefont {Y.}~\bibnamefont
			{Wang}}, \bibinfo {author} {\bibfnamefont {Q.}~\bibnamefont {Chen}}, \bibinfo
		{author} {\bibfnamefont {W.}~\bibnamefont {Yang}}, \bibinfo {author}
		{\bibfnamefont {Z.}~\bibnamefont {Ji}}, \bibinfo {author} {\bibfnamefont
			{L.}~\bibnamefont {Jin}}, \bibinfo {author} {\bibfnamefont {X.}~\bibnamefont
			{Ma}}, \bibinfo {author} {\bibfnamefont {Q.}~\bibnamefont {Song}}, \bibinfo
		{author} {\bibfnamefont {A.}~\bibnamefont {Boltasseva}}, \bibinfo {author}
		{\bibfnamefont {J.}~\bibnamefont {Han}}, \bibinfo {author} {\bibfnamefont
			{V.~M.}\ \bibnamefont {Shalaev}},\ and\ \bibinfo {author} {\bibfnamefont
			{S.}~\bibnamefont {Xiao}},\ }\bibfield  {title} {\enquote {\bibinfo {title}
			{High-efficiency broadband achromatic metalens for near-ir biological imaging
				window},}\ }\href {https://doi.org/10.1038/s41467-021-25797-9} {\bibfield
		{journal} {\bibinfo  {journal} {Nature Communications}\ }\textbf {\bibinfo
			{volume} {12}},\ \bibinfo {pages} {5560} (\bibinfo {year}
		{2021})}\BibitemShut {NoStop}%
	\bibitem [{\citenamefont {Hu}\ \emph {et~al.}(2023)\citenamefont {Hu},
		\citenamefont {Jiang}, \citenamefont {Zhang}, \citenamefont {Yang},
		\citenamefont {Ou}, \citenamefont {Li}, \citenamefont {Kong}, \citenamefont
		{Liu}, \citenamefont {Qiu},\ and\ \citenamefont {Duan}}]{Hu2023}%
	\BibitemOpen
	\bibfield  {author} {\bibinfo {author} {\bibfnamefont {Y.}~\bibnamefont
			{Hu}}, \bibinfo {author} {\bibfnamefont {Y.}~\bibnamefont {Jiang}}, \bibinfo
		{author} {\bibfnamefont {Y.}~\bibnamefont {Zhang}}, \bibinfo {author}
		{\bibfnamefont {X.}~\bibnamefont {Yang}}, \bibinfo {author} {\bibfnamefont
			{X.}~\bibnamefont {Ou}}, \bibinfo {author} {\bibfnamefont {L.}~\bibnamefont
			{Li}}, \bibinfo {author} {\bibfnamefont {X.}~\bibnamefont {Kong}}, \bibinfo
		{author} {\bibfnamefont {X.}~\bibnamefont {Liu}}, \bibinfo {author}
		{\bibfnamefont {C.-W.}\ \bibnamefont {Qiu}},\ and\ \bibinfo {author}
		{\bibfnamefont {H.}~\bibnamefont {Duan}},\ }\bibfield  {title} {\enquote
		{\bibinfo {title} {Asymptotic dispersion engineering for ultra-broadband
				meta-optics},}\ }\href {https://doi.org/10.1038/s41467-023-42268-5}
	{\bibfield  {journal} {\bibinfo  {journal} {Nature Communications}\ }\textbf
		{\bibinfo {volume} {14}},\ \bibinfo {pages} {6649} (\bibinfo {year}
		{2023})}\BibitemShut {NoStop}%
	\bibitem [{\citenamefont {Yoon}\ \emph {et~al.}(2021)\citenamefont {Yoon},
		\citenamefont {Kim}, \citenamefont {Kim}, \citenamefont {Han}, \citenamefont
		{Lee},\ and\ \citenamefont {Rho}}]{Yoon2021}%
	\BibitemOpen
	\bibfield  {author} {\bibinfo {author} {\bibfnamefont {G.}~\bibnamefont
			{Yoon}}, \bibinfo {author} {\bibfnamefont {K.}~\bibnamefont {Kim}}, \bibinfo
		{author} {\bibfnamefont {S.-U.}\ \bibnamefont {Kim}}, \bibinfo {author}
		{\bibfnamefont {S.}~\bibnamefont {Han}}, \bibinfo {author} {\bibfnamefont
			{H.}~\bibnamefont {Lee}},\ and\ \bibinfo {author} {\bibfnamefont
			{J.}~\bibnamefont {Rho}},\ }\bibfield  {title} {\enquote {\bibinfo {title}
			{Printable nanocomposite metalens for high-contrast near-infrared imaging},}\
	}\href {https://doi.org/10.1021/acsnano.0c06968} {\bibfield  {journal}
		{\bibinfo  {journal} {ACS Nano}\ }\textbf {\bibinfo {volume} {15}},\ \bibinfo
		{pages} {698--706} (\bibinfo {year} {2021})}\BibitemShut {NoStop}%
	\bibitem [{\citenamefont {Li}\ \emph {et~al.}(2021{\natexlab{b}})\citenamefont
		{Li}, \citenamefont {Xu}, \citenamefont {Fu}, \citenamefont {Wang},
		\citenamefont {Li}, \citenamefont {Wang},\ and\ \citenamefont
		{Zhu}}]{Li2021a}%
	\BibitemOpen
	\bibfield  {author} {\bibinfo {author} {\bibfnamefont {T.}~\bibnamefont
			{Li}}, \bibinfo {author} {\bibfnamefont {X.}~\bibnamefont {Xu}}, \bibinfo
		{author} {\bibfnamefont {B.}~\bibnamefont {Fu}}, \bibinfo {author}
		{\bibfnamefont {S.}~\bibnamefont {Wang}}, \bibinfo {author} {\bibfnamefont
			{B.}~\bibnamefont {Li}}, \bibinfo {author} {\bibfnamefont {Z.}~\bibnamefont
			{Wang}},\ and\ \bibinfo {author} {\bibfnamefont {S.}~\bibnamefont {Zhu}},\
	}\bibfield  {title} {\enquote {\bibinfo {title} {Integrating the optical
				tweezers and spanner onto an individual single-layer metasurface},}\ }\href
	{https://doi.org/10.1364/prj.421121} {\bibfield  {journal} {\bibinfo
			{journal} {Photonics Research}\ }\textbf {\bibinfo {volume} {9}},\ \bibinfo
		{pages} {1062} (\bibinfo {year} {2021}{\natexlab{b}})}\BibitemShut {NoStop}%
	\bibitem [{\citenamefont {Li}\ \emph {et~al.}(2020)\citenamefont {Li},
		\citenamefont {Liu}, \citenamefont {Ren}, \citenamefont {Wang}, \citenamefont
		{Su}, \citenamefont {Chen}, \citenamefont {Chu}, \citenamefont {Kuo},
		\citenamefont {Liu}, \citenamefont {Zang}, \citenamefont {Guo}, \citenamefont
		{Zhang}, \citenamefont {Wang}, \citenamefont {Zhu},\ and\ \citenamefont
		{Tsai}}]{Li2020}%
	\BibitemOpen
	\bibfield  {author} {\bibinfo {author} {\bibfnamefont {L.}~\bibnamefont
			{Li}}, \bibinfo {author} {\bibfnamefont {Z.}~\bibnamefont {Liu}}, \bibinfo
		{author} {\bibfnamefont {X.}~\bibnamefont {Ren}}, \bibinfo {author}
		{\bibfnamefont {S.}~\bibnamefont {Wang}}, \bibinfo {author} {\bibfnamefont
			{V.-C.}\ \bibnamefont {Su}}, \bibinfo {author} {\bibfnamefont {M.-K.}\
			\bibnamefont {Chen}}, \bibinfo {author} {\bibfnamefont {C.~H.}\ \bibnamefont
			{Chu}}, \bibinfo {author} {\bibfnamefont {H.~Y.}\ \bibnamefont {Kuo}},
		\bibinfo {author} {\bibfnamefont {B.}~\bibnamefont {Liu}}, \bibinfo {author}
		{\bibfnamefont {W.}~\bibnamefont {Zang}}, \bibinfo {author} {\bibfnamefont
			{G.}~\bibnamefont {Guo}}, \bibinfo {author} {\bibfnamefont {L.}~\bibnamefont
			{Zhang}}, \bibinfo {author} {\bibfnamefont {Z.}~\bibnamefont {Wang}},
		\bibinfo {author} {\bibfnamefont {S.}~\bibnamefont {Zhu}},\ and\ \bibinfo
		{author} {\bibfnamefont {D.~P.}\ \bibnamefont {Tsai}},\ }\bibfield  {title}
	{\enquote {\bibinfo {title} {Metalens-array-based high-dimensional and
				multiphoton quantum source},}\ }\href
	{https://doi.org/10.1126/science.aba9779} {\bibfield  {journal} {\bibinfo
			{journal} {Science}\ }\textbf {\bibinfo {volume} {368}},\ \bibinfo {pages}
		{1487--1490} (\bibinfo {year} {2020})}\BibitemShut {NoStop}%
	\bibitem [{\citenamefont {Li}\ \emph {et~al.}(2023)\citenamefont {Li},
		\citenamefont {Jang}, \citenamefont {Badloe}, \citenamefont {Yang},
		\citenamefont {Kim}, \citenamefont {Kim}, \citenamefont {Nguyen},
		\citenamefont {Maier}, \citenamefont {Rho}, \citenamefont {Ren},\ and\
		\citenamefont {Aharonovich}}]{Li2023}%
	\BibitemOpen
	\bibfield  {author} {\bibinfo {author} {\bibfnamefont {C.}~\bibnamefont
			{Li}}, \bibinfo {author} {\bibfnamefont {J.}~\bibnamefont {Jang}}, \bibinfo
		{author} {\bibfnamefont {T.}~\bibnamefont {Badloe}}, \bibinfo {author}
		{\bibfnamefont {T.}~\bibnamefont {Yang}}, \bibinfo {author} {\bibfnamefont
			{J.}~\bibnamefont {Kim}}, \bibinfo {author} {\bibfnamefont {J.}~\bibnamefont
			{Kim}}, \bibinfo {author} {\bibfnamefont {M.}~\bibnamefont {Nguyen}},
		\bibinfo {author} {\bibfnamefont {S.~A.}\ \bibnamefont {Maier}}, \bibinfo
		{author} {\bibfnamefont {J.}~\bibnamefont {Rho}}, \bibinfo {author}
		{\bibfnamefont {H.}~\bibnamefont {Ren}},\ and\ \bibinfo {author}
		{\bibfnamefont {I.}~\bibnamefont {Aharonovich}},\ }\bibfield  {title}
	{\enquote {\bibinfo {title} {Arbitrarily structured quantum emission with a
				multifunctional metalens},}\ }\href
	{https://doi.org/10.1186/s43593-023-00052-4} {\bibfield  {journal} {\bibinfo
			{journal} {eLight}\ }\textbf {\bibinfo {volume} {3}},\ \bibinfo {pages} {19}
		(\bibinfo {year} {2023})}\BibitemShut {NoStop}%
	\bibitem [{\citenamefont {Xiao}\ \emph {et~al.}(2020)\citenamefont {Xiao},
		\citenamefont {Wang}, \citenamefont {Liu}, \citenamefont {Zhou},
		\citenamefont {Jiang},\ and\ \citenamefont {Zhang}}]{Xiao2020}%
	\BibitemOpen
	\bibfield  {author} {\bibinfo {author} {\bibfnamefont {S.}~\bibnamefont
			{Xiao}}, \bibinfo {author} {\bibfnamefont {T.}~\bibnamefont {Wang}}, \bibinfo
		{author} {\bibfnamefont {T.}~\bibnamefont {Liu}}, \bibinfo {author}
		{\bibfnamefont {C.}~\bibnamefont {Zhou}}, \bibinfo {author} {\bibfnamefont
			{X.}~\bibnamefont {Jiang}},\ and\ \bibinfo {author} {\bibfnamefont
			{J.}~\bibnamefont {Zhang}},\ }\bibfield  {title} {\enquote {\bibinfo {title}
			{Active metamaterials and metadevices: a review},}\ }\href
	{https://doi.org/10.1088/1361-6463/abaced} {\bibfield  {journal} {\bibinfo
			{journal} {Journal of Physics D: Applied Physics}\ }\textbf {\bibinfo
			{volume} {53}},\ \bibinfo {pages} {503002} (\bibinfo {year}
		{2020})}\BibitemShut {NoStop}%
	\bibitem [{\citenamefont {Ee}\ and\ \citenamefont {Agarwal}(2016)}]{Ee2016}%
	\BibitemOpen
	\bibfield  {author} {\bibinfo {author} {\bibfnamefont {H.-S.}\ \bibnamefont
			{Ee}}\ and\ \bibinfo {author} {\bibfnamefont {R.}~\bibnamefont {Agarwal}},\
	}\bibfield  {title} {\enquote {\bibinfo {title} {Tunable metasurface and flat
				optical zoom lens on a stretchable substrate},}\ }\href
	{https://doi.org/10.1021/acs.nanolett.6b00618} {\bibfield  {journal}
		{\bibinfo  {journal} {Nano Letters}\ }\textbf {\bibinfo {volume} {16}},\
		\bibinfo {pages} {2818--2823} (\bibinfo {year} {2016})}\BibitemShut {NoStop}%
	\bibitem [{\citenamefont {Bosch}\ \emph {et~al.}(2021)\citenamefont {Bosch},
		\citenamefont {Shcherbakov}, \citenamefont {Won}, \citenamefont {Lee},
		\citenamefont {Kim},\ and\ \citenamefont {Shvets}}]{Bosch2021}%
	\BibitemOpen
	\bibfield  {author} {\bibinfo {author} {\bibfnamefont {M.}~\bibnamefont
			{Bosch}}, \bibinfo {author} {\bibfnamefont {M.~R.}\ \bibnamefont
			{Shcherbakov}}, \bibinfo {author} {\bibfnamefont {K.}~\bibnamefont {Won}},
		\bibinfo {author} {\bibfnamefont {H.-S.}\ \bibnamefont {Lee}}, \bibinfo
		{author} {\bibfnamefont {Y.}~\bibnamefont {Kim}},\ and\ \bibinfo {author}
		{\bibfnamefont {G.}~\bibnamefont {Shvets}},\ }\bibfield  {title} {\enquote
		{\bibinfo {title} {Electrically actuated varifocal lens based on
				liquid-crystal-embedded dielectric metasurfaces},}\ }\href
	{https://doi.org/10.1021/acs.nanolett.1c00356} {\bibfield  {journal}
		{\bibinfo  {journal} {Nano Letters}\ }\textbf {\bibinfo {volume} {21}},\
		\bibinfo {pages} {3849--3856} (\bibinfo {year} {2021})}\BibitemShut {NoStop}%
	\bibitem [{\citenamefont {Zhu}\ \emph {et~al.}(2023)\citenamefont {Zhu},
		\citenamefont {Jiang}, \citenamefont {Wang},\ and\ \citenamefont
		{Huang}}]{Zhu2023}%
	\BibitemOpen
	\bibfield  {author} {\bibinfo {author} {\bibfnamefont {S.}~\bibnamefont
			{Zhu}}, \bibinfo {author} {\bibfnamefont {Q.}~\bibnamefont {Jiang}}, \bibinfo
		{author} {\bibfnamefont {Y.}~\bibnamefont {Wang}},\ and\ \bibinfo {author}
		{\bibfnamefont {L.}~\bibnamefont {Huang}},\ }\bibfield  {title} {\enquote
		{\bibinfo {title} {Nonmechanical varifocal metalens using nematic liquid
				crystal},}\ }\href {https://doi.org/10.1515/nanoph-2023-0001} {\bibfield
		{journal} {\bibinfo  {journal} {Nanophotonics}\ }\textbf {\bibinfo {volume}
			{12}},\ \bibinfo {pages} {1169--1176} (\bibinfo {year} {2023})}\BibitemShut
	{NoStop}%
	\bibitem [{\citenamefont {Luo}\ \emph {et~al.}(2021)\citenamefont {Luo},
		\citenamefont {Chu}, \citenamefont {Vyas}, \citenamefont {Kuo}, \citenamefont
		{Chia}, \citenamefont {Chen}, \citenamefont {Shi}, \citenamefont {Tanaka},
		\citenamefont {Misawa}, \citenamefont {Huang},\ and\ \citenamefont
		{Tsai}}]{Luo2021}%
	\BibitemOpen
	\bibfield  {author} {\bibinfo {author} {\bibfnamefont {Y.}~\bibnamefont
			{Luo}}, \bibinfo {author} {\bibfnamefont {C.~H.}\ \bibnamefont {Chu}},
		\bibinfo {author} {\bibfnamefont {S.}~\bibnamefont {Vyas}}, \bibinfo {author}
		{\bibfnamefont {H.~Y.}\ \bibnamefont {Kuo}}, \bibinfo {author} {\bibfnamefont
			{Y.~H.}\ \bibnamefont {Chia}}, \bibinfo {author} {\bibfnamefont {M.~K.}\
			\bibnamefont {Chen}}, \bibinfo {author} {\bibfnamefont {X.}~\bibnamefont
			{Shi}}, \bibinfo {author} {\bibfnamefont {T.}~\bibnamefont {Tanaka}},
		\bibinfo {author} {\bibfnamefont {H.}~\bibnamefont {Misawa}}, \bibinfo
		{author} {\bibfnamefont {Y.-Y.}\ \bibnamefont {Huang}},\ and\ \bibinfo
		{author} {\bibfnamefont {D.~P.}\ \bibnamefont {Tsai}},\ }\bibfield  {title}
	{\enquote {\bibinfo {title} {Varifocal metalens for optical sectioning
				fluorescence microscopy},}\ }\href
	{https://doi.org/10.1021/acs.nanolett.1c01114} {\bibfield  {journal}
		{\bibinfo  {journal} {Nano Letters}\ }\textbf {\bibinfo {volume} {21}},\
		\bibinfo {pages} {5133--5142} (\bibinfo {year} {2021})}\BibitemShut {NoStop}%
	\bibitem [{\citenamefont {Ding}, \citenamefont {Yang},\ and\ \citenamefont
		{Bozhevolnyi}(2019)}]{Ding2019}%
	\BibitemOpen
	\bibfield  {author} {\bibinfo {author} {\bibfnamefont {F.}~\bibnamefont
			{Ding}}, \bibinfo {author} {\bibfnamefont {Y.}~\bibnamefont {Yang}},\ and\
		\bibinfo {author} {\bibfnamefont {S.~I.}\ \bibnamefont {Bozhevolnyi}},\
	}\bibfield  {title} {\enquote {\bibinfo {title} {Dynamic metasurfaces using
				phase-change chalcogenides},}\ }\href
	{https://doi.org/10.1002/adom.201801709} {\bibfield  {journal} {\bibinfo
			{journal} {Advanced Optical Materials}\ }\textbf {\bibinfo {volume} {7}},\
		\bibinfo {pages} {1801709} (\bibinfo {year} {2019})}\BibitemShut {NoStop}%
	\bibitem [{\citenamefont {Qin}\ \emph {et~al.}(2021)\citenamefont {Qin},
		\citenamefont {Xu}, \citenamefont {Huang}, \citenamefont {Jie}, \citenamefont
		{Liu}, \citenamefont {Guo}, \citenamefont {Meng}, \citenamefont {Wang},
		\citenamefont {Yang},\ and\ \citenamefont {Wei}}]{Qin2021}%
	\BibitemOpen
	\bibfield  {author} {\bibinfo {author} {\bibfnamefont {S.}~\bibnamefont
			{Qin}}, \bibinfo {author} {\bibfnamefont {N.}~\bibnamefont {Xu}}, \bibinfo
		{author} {\bibfnamefont {H.}~\bibnamefont {Huang}}, \bibinfo {author}
		{\bibfnamefont {K.}~\bibnamefont {Jie}}, \bibinfo {author} {\bibfnamefont
			{H.}~\bibnamefont {Liu}}, \bibinfo {author} {\bibfnamefont {J.}~\bibnamefont
			{Guo}}, \bibinfo {author} {\bibfnamefont {H.}~\bibnamefont {Meng}}, \bibinfo
		{author} {\bibfnamefont {F.}~\bibnamefont {Wang}}, \bibinfo {author}
		{\bibfnamefont {X.}~\bibnamefont {Yang}},\ and\ \bibinfo {author}
		{\bibfnamefont {Z.}~\bibnamefont {Wei}},\ }\bibfield  {title} {\enquote
		{\bibinfo {title} {Near-infrared thermally modulated varifocal metalens based
				on the phase change material sb2s3},}\ }\href
	{https://doi.org/10.1364/oe.420014} {\bibfield  {journal} {\bibinfo
			{journal} {Optics Express}\ }\textbf {\bibinfo {volume} {29}},\ \bibinfo
		{pages} {7925} (\bibinfo {year} {2021})}\BibitemShut {NoStop}%
	\bibitem [{\citenamefont {Li}\ \emph {et~al.}(2024{\natexlab{a}})\citenamefont
		{Li}, \citenamefont {Guo}, \citenamefont {Wang}, \citenamefont {Lu},
		\citenamefont {Qiu}, \citenamefont {Hou},\ and\ \citenamefont
		{Guo}}]{Li2024}%
	\BibitemOpen
	\bibfield  {author} {\bibinfo {author} {\bibfnamefont {Z.}~\bibnamefont
			{Li}}, \bibinfo {author} {\bibfnamefont {P.}~\bibnamefont {Guo}}, \bibinfo
		{author} {\bibfnamefont {Z.}~\bibnamefont {Wang}}, \bibinfo {author}
		{\bibfnamefont {B.}~\bibnamefont {Lu}}, \bibinfo {author} {\bibfnamefont
			{W.}~\bibnamefont {Qiu}}, \bibinfo {author} {\bibfnamefont {W.}~\bibnamefont
			{Hou}},\ and\ \bibinfo {author} {\bibfnamefont {L.}~\bibnamefont {Guo}},\
	}\bibfield  {title} {\enquote {\bibinfo {title} {Design of a high-efficiency
				reconfigurable anomalous transmission metasurface using sb2se3},}\ }\href
	{https://doi.org/10.1109/lpt.2024.3409761} {\bibfield  {journal} {\bibinfo
			{journal} {IEEE Photonics Technology Letters}\ }\textbf {\bibinfo {volume}
			{36}},\ \bibinfo {pages} {905--908} (\bibinfo {year}
		{2024}{\natexlab{a}})}\BibitemShut {NoStop}%
	\bibitem [{\citenamefont {Zhang}, \citenamefont {Yao},\ and\ \citenamefont
		{Tsai}(2024)}]{Zhang2024}%
	\BibitemOpen
	\bibfield  {author} {\bibinfo {author} {\bibfnamefont {J.~C.}\ \bibnamefont
			{Zhang}}, \bibinfo {author} {\bibfnamefont {J.}~\bibnamefont {Yao}},\ and\
		\bibinfo {author} {\bibfnamefont {D.~P.}\ \bibnamefont {Tsai}},\ }\bibfield
	{title} {\enquote {\bibinfo {title} {Meta-lens based on multi-level
				phase-change},}\ }\href {https://doi.org/10.1063/5.0221280} {\bibfield
		{journal} {\bibinfo  {journal} {Journal of Applied Physics}\ }\textbf
		{\bibinfo {volume} {136}},\ \bibinfo {pages} {053101} (\bibinfo {year}
		{2024})}\BibitemShut {NoStop}%
	\bibitem [{\citenamefont {Zhang}\ \emph {et~al.}(2023)\citenamefont {Zhang},
		\citenamefont {Fang}, \citenamefont {Zhao}, \citenamefont {Li}, \citenamefont
		{Shen}, \citenamefont {Hong},\ and\ \citenamefont {Jing}}]{Zhang2023}%
	\BibitemOpen
	\bibfield  {author} {\bibinfo {author} {\bibfnamefont {P.}~\bibnamefont
			{Zhang}}, \bibinfo {author} {\bibfnamefont {B.}~\bibnamefont {Fang}},
		\bibinfo {author} {\bibfnamefont {T.}~\bibnamefont {Zhao}}, \bibinfo {author}
		{\bibfnamefont {C.}~\bibnamefont {Li}}, \bibinfo {author} {\bibfnamefont
			{C.}~\bibnamefont {Shen}}, \bibinfo {author} {\bibfnamefont {Z.}~\bibnamefont
			{Hong}},\ and\ \bibinfo {author} {\bibfnamefont {X.}~\bibnamefont {Jing}},\
	}\bibfield  {title} {\enquote {\bibinfo {title} {Terahertz wave tunable
				metalens based on phase change material coded metasurface},}\ }\href
	{https://doi.org/10.1109/jlt.2023.3262509} {\bibfield  {journal} {\bibinfo
			{journal} {Journal of Lightwave Technology}\ }\textbf {\bibinfo {volume}
			{41}},\ \bibinfo {pages} {7162--7168} (\bibinfo {year} {2023})}\BibitemShut
	{NoStop}%
	\bibitem [{\citenamefont {Lu}\ \emph {et~al.}(2021)\citenamefont {Lu},
		\citenamefont {Dong}, \citenamefont {Tijiptoharsono}, \citenamefont {Ng},
		\citenamefont {Wang}, \citenamefont {Rezaei}, \citenamefont {Wang},
		\citenamefont {Leong}, \citenamefont {Lim}, \citenamefont {Yang},\ and\
		\citenamefont {Simpson}}]{Lu2021}%
	\BibitemOpen
	\bibfield  {author} {\bibinfo {author} {\bibfnamefont {L.}~\bibnamefont
			{Lu}}, \bibinfo {author} {\bibfnamefont {Z.}~\bibnamefont {Dong}}, \bibinfo
		{author} {\bibfnamefont {F.}~\bibnamefont {Tijiptoharsono}}, \bibinfo
		{author} {\bibfnamefont {R.~J.~H.}\ \bibnamefont {Ng}}, \bibinfo {author}
		{\bibfnamefont {H.}~\bibnamefont {Wang}}, \bibinfo {author} {\bibfnamefont
			{S.~D.}\ \bibnamefont {Rezaei}}, \bibinfo {author} {\bibfnamefont
			{Y.}~\bibnamefont {Wang}}, \bibinfo {author} {\bibfnamefont {H.~S.}\
			\bibnamefont {Leong}}, \bibinfo {author} {\bibfnamefont {P.~C.}\ \bibnamefont
			{Lim}}, \bibinfo {author} {\bibfnamefont {J.~K.~W.}\ \bibnamefont {Yang}},\
		and\ \bibinfo {author} {\bibfnamefont {R.~E.}\ \bibnamefont {Simpson}},\
	}\bibfield  {title} {\enquote {\bibinfo {title} {Reversible tuning of mie
				resonances in the visible spectrum},}\ }\href
	{https://doi.org/10.1021/acsnano.1c07114} {\bibfield  {journal} {\bibinfo
			{journal} {ACS Nano}\ }\textbf {\bibinfo {volume} {15}},\ \bibinfo {pages}
		{19722--19732} (\bibinfo {year} {2021})}\BibitemShut {NoStop}%
	\bibitem [{\citenamefont {Li}\ \emph {et~al.}(2024{\natexlab{b}})\citenamefont
		{Li}, \citenamefont {Lu}, \citenamefont {Li}, \citenamefont {Song},
		\citenamefont {Tan}, \citenamefont {He}, \citenamefont {Liu}, \citenamefont
		{Luo}, \citenamefont {Tang}, \citenamefont {Liu}, \citenamefont {Xu},
		\citenamefont {Xiao}, \citenamefont {Huang}, \citenamefont {Shen},
		\citenamefont {Zhang}, \citenamefont {Zhang},\ and\ \citenamefont
		{Yao}}]{Li2024a}%
	\BibitemOpen
	\bibfield  {author} {\bibinfo {author} {\bibfnamefont {J.}~\bibnamefont
			{Li}}, \bibinfo {author} {\bibfnamefont {X.}~\bibnamefont {Lu}}, \bibinfo
		{author} {\bibfnamefont {H.}~\bibnamefont {Li}}, \bibinfo {author}
		{\bibfnamefont {C.}~\bibnamefont {Song}}, \bibinfo {author} {\bibfnamefont
			{Q.}~\bibnamefont {Tan}}, \bibinfo {author} {\bibfnamefont {Y.}~\bibnamefont
			{He}}, \bibinfo {author} {\bibfnamefont {J.}~\bibnamefont {Liu}}, \bibinfo
		{author} {\bibfnamefont {L.}~\bibnamefont {Luo}}, \bibinfo {author}
		{\bibfnamefont {T.}~\bibnamefont {Tang}}, \bibinfo {author} {\bibfnamefont
			{T.}~\bibnamefont {Liu}}, \bibinfo {author} {\bibfnamefont {H.}~\bibnamefont
			{Xu}}, \bibinfo {author} {\bibfnamefont {S.}~\bibnamefont {Xiao}}, \bibinfo
		{author} {\bibfnamefont {W.}~\bibnamefont {Huang}}, \bibinfo {author}
		{\bibfnamefont {Y.}~\bibnamefont {Shen}}, \bibinfo {author} {\bibfnamefont
			{Y.}~\bibnamefont {Zhang}}, \bibinfo {author} {\bibfnamefont
			{Y.}~\bibnamefont {Zhang}},\ and\ \bibinfo {author} {\bibfnamefont
			{J.}~\bibnamefont {Yao}},\ }\bibfield  {title} {\enquote {\bibinfo {title}
			{Racemic dielectric metasurfaces for arbitrary terahertz polarization
				rotation and wavefront manipulation},}\ }\href
	{https://doi.org/10.29026/oea.2024.240075} {\bibfield  {journal} {\bibinfo
			{journal} {Opto-Electronic Advances}\ }\textbf {\bibinfo {volume} {7}},\
		\bibinfo {pages} {240075} (\bibinfo {year} {2024}{\natexlab{b}})}\BibitemShut
	{NoStop}%
\end{thebibliography}

%

\end{document}